\documentclass[twocolumn,showpacs,preprintnumbers,amsmath,amssymb]{revtex4}

\makeatletter

\usepackage{graphicx}
\usepackage{dcolumn}
\usepackage{bm}

\begin{document}

\title{Propagation of Light in Cantor Media}

\author{Masanori Yamanaka and Mahito Kohmoto$^1$}
\affiliation{Department of Physics, College of Science and Technology,
Nihon University, Kanda-Surugadai 1-8-14, Chiyoda-ku,
Tokyo 101-8308, Japan}
\affiliation{
$^1$Institute for Solid State Physics, University of Tokyo,
Kashiwanoha 5-1-5, Kashiwa, Chiba, 277-8581, Japan
}

\date{\today}

\begin{abstract}
We numerically find that transmission coefficients 
have a rich structure as a function of wavelength 
in Cantor media.
Complete transmission and complete reflection are observed.
We also find that light propagation has scalings
with respect to number of layers.
\end{abstract}

\pacs{42.25.-p, 71.55.Jv, 46.65.+g, 61.44.-n}

\maketitle

Localization of electronic states due to disorder 
is one of the most active fields 
in condensed-matter physics~\cite{REF:localization}.
It has been widely recognized that
localization could occur not only in disordered systems
but also in the quasiperiodic systems
in one dimension~\cite{REF:quasiperiodic}.
In a quasiperiodic system two (or more) incommensurate periods 
are superposed, so that it is neither a periodic nor a random system
and could be considered to be intermediate between the two.

In one dimension, a quasiperiodic Schr\"odinger equation
based on the Fibonacci sequence has been analyzed 
by a renormalization-group type theory~\cite{REF:fibonacci1,REF:fibonacci2}.
In this model, a simple binary quasiperiodic sequence 
is used which is constructed recursively 
as $F_{j+1}=\{F_{j-1}, F_j\}$, for $j\ge1$,
with $F_0=\{ B \}$ and $F_1=\{ A \}$.
In this sequence one has $F_2=\{ BA \}$, $F_3=\{ ABA \}$, 
$F_4=\{ BAABA \}$, and so on.
The most striking feature of this model 
is that all the states are critical.
Namely, the wave functions are not localized exponentially
but only weakly localized and have a rich structure 
including scaling~\cite{REF:fibonacci2,REF:KST}.
The energy spectrum also has a rich structure.
It was found to have zero Lebesgue measure. 
Namely, if one picks an energy, it is in a gap 
with probability one and the gaps are dense.
Also there are no isolated points.
The spectrum has a self-similar structure 
with various scaling indices 
(multifractal)~\cite{REF:KST,REFsuto}.

While the localization of states was originally regarded
as an electronic problem, it was later recognized 
that the phenomenon is essentially a consequence 
of the wave nature of the electronic states.
Therefore, such localization can be expected 
for any wave phenomenon.
An optical experiment with the Fibonacci layer 
was proposed~\cite{REF:KSI}.
The corresponding experiments 
were reported~\cite{REF:GKST,REF:HTKN}
using a dielectric multilayer stacks 
of SiO$_2$ and TiO$_2$ thin films.
See also~\cite{REF:Chow}.
The transmission coefficients and the scaling properties
remarkably agree with the theory~\cite{REF:KSI}
and can be considered as an experimental evidence 
for localization of light waves.
Recently, an experiment for propagation 
of electromagnetic wave 
in a three-dimensional fractal cavity was reported~\cite{REF:3Dfractal}.
The material is called the Menger sponge
and the structure is the cubic Sierpinski gasket
which has a self-similar pattern 
with single scaling index (not multifractal).
The results are interpreted as a localization 
of the electromagnetic waves~\cite{REF:3Dfractal}.

\noindent
\begin{figure}[b]
\includegraphics[width=8cm]{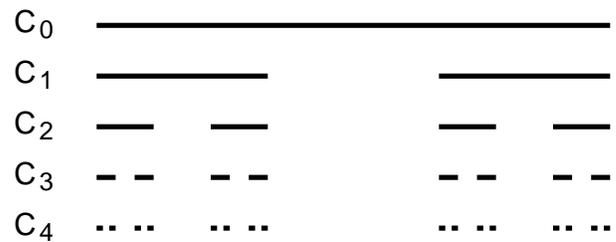}
\caption{
The initial condition $j=0$ and the first four 
generations in the construction of the Cantor set.
The first generation, i.e. $n=1$, provides a ``generator.'' 
We call the $j$-th generation a Cantor sequence $C_j$. 
}
\label{FIG:Cantor}
\end{figure}

In this paper, we propose an optical experiment with Cantor layers.
In this system one-dimensional theory is strictly valid.
Also, it is feasible to construct a system accurately
and the parameters are precisely controlled and measured.
We calculate the transmission coefficient as a function 
of wavelength of light.
The results show a singular structure, 
i.e. it alternates between complete transmission 
and complete reflection. 
The presence of complete reflection is quite striking,
because the substrate of the Cantor layer has zero Lebesgue measure 
in the thermodynamic limit of the generation.
Namely, if one picks the substrate, it is in vacuum
with probability one and the vacuum is dense.
Also there are no isolated points.
We also find that the light propagation has scaling 
with respect to the number of layers.

The procedure of constructing the Cantor set begins 
with a line segment of unit length. 
We regard this as substrate $A$.
This line segment is divided into three equal parts
and the middle part is removed to obtain the first generation;
that serves as the ``generator'' of the Cantor set.
The removed area is substrate $B$.
The procedure is repeated for each of the two line segments
of the first generation to obtain the second generation and so on.
Therefore the $j$-th generation of the Cantor set is a finite
set of $2^j$ line segments, each of length $1/3^j$.
If this procedure is repeated an infinite number of times
the remainder set of discrete points is called the Cantor set.
It is an exact self-similar fractal of dimension $\log_32$,
which is a single scaling dimension and is not a multi-fractal. 
The construction of first few generation is shown 
in Fig.\ref{FIG:Cantor}.

\noindent
\begin{figure}[b]
\includegraphics[width=6.5cm]{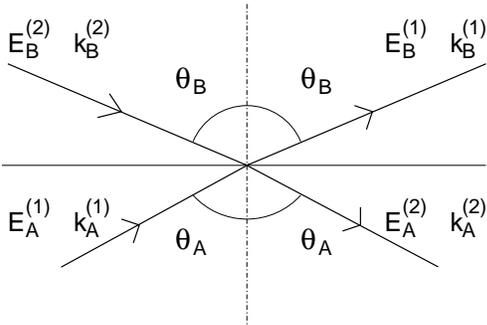}
\caption{
Electromagnetic wave propagation across an interface 
of two layers $A$ and $B$.
}
\label{FIG:waveprop}
\end{figure}

\noindent
\begin{figure}[b]
\includegraphics[width=8.5cm]{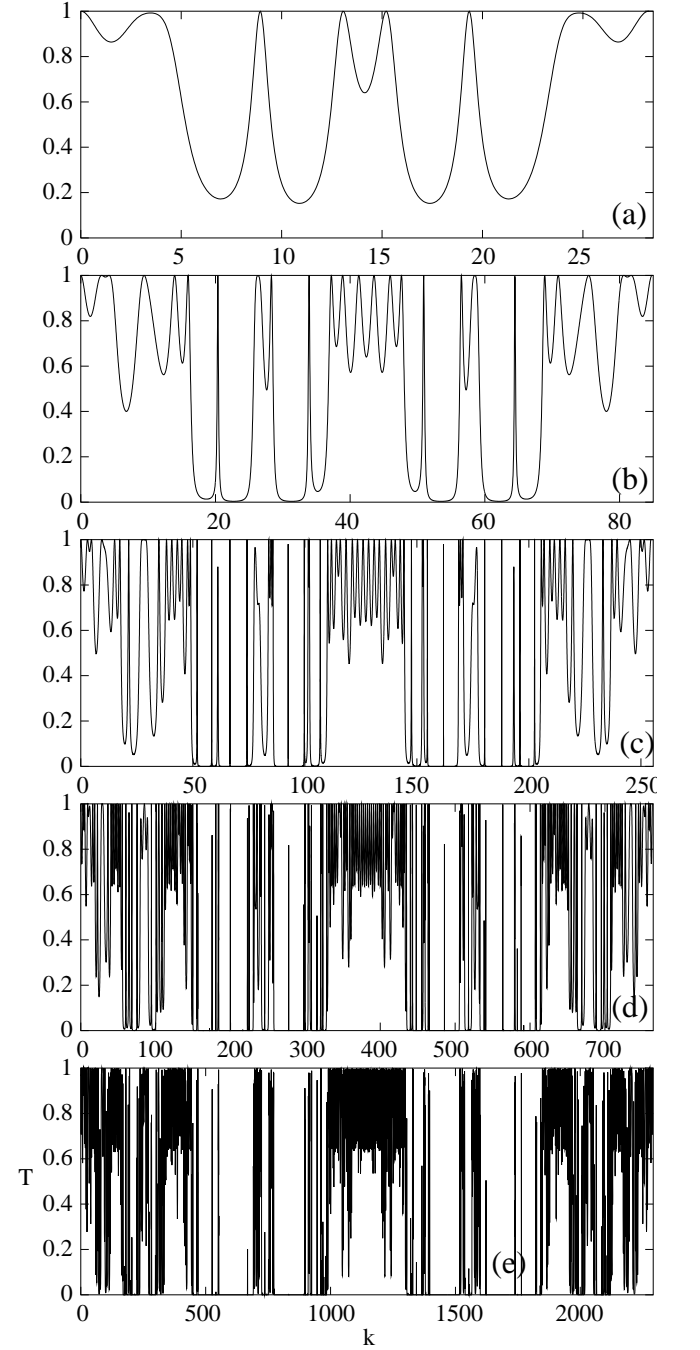}
\caption{
Transmission coefficient $T$ as a function 
of the wave number $k$ 
for multilayers $C_2$ (a), $C_3$ (b), $C_4$ (c),
$C_5$ (d), and $C_6$ (e).
}
\label{FIG:transn=2}
\end{figure}

Let us consider a multilayer in which two types of layers
$A$ and $B$ are arranged in a Cantor sequence.
In order to understand the light propagation in this media,
first consider an interface of two layers.
See Fig.\ref{FIG:waveprop}.
The electric field for the light in layer $A$ is given by 
\begin{eqnarray} 
\vec{E}=\vec{E}_A^{(1)}
 e^{i(\vec{k}_A^{(1)} \cdot \vec{x}-\omega t)}
+\vec{E}_A^{(2)}
 e^{i(\vec{k}_A^{(2)} \cdot \vec{x}-\omega t)}.
\label{eq:electric}
\end{eqnarray} 
The electric field in layer $B$ is given by the same expression
with subscript $A$ replaced by $B$.
We consider a polarization which is perpendicular 
to the plane of the light path (TE wave).
The appropriate boundary condition at an interface gives
\begin{eqnarray} 
&&E_A^{(1)}+E_A^{(2)}=E_B^{(1)}+E_B^{(2)},
\nonumber\\
&&n_A \cos \theta_A (E_A^{(1)}-E_A^{(2)})
 =n_B \cos \theta_B (E_B^{(1)}-E_B^{(2)}),
\label{eq:boundary}
\end{eqnarray} 
where $n_A$ and $n_B$ are indices of refraction of $A$ and $B$,
respectively, and the angles $\theta_A$ and $\theta_B$ are shown
in Fig.\ref{FIG:waveprop}.
Snell's law is $\sin\theta_A/\sin\theta_B=n_B/n_A$.
It is convenient to choose the two independent variables
for the light as
\begin{eqnarray} 
E_+=E^{(1)}+E^{(2)}, \ \ \ E_-=(E^{(1)}-E^{(2)})/i.
\label{eq:redef}
\end{eqnarray} 
Then (\ref{eq:boundary}) gives
\begin{eqnarray} 
\left[
\begin{array}{c}
E_+ \\
E_- 
\end{array}
\right]_B
= T_{BA}
\left[
\begin{array}{c}
E_+ \\
E_- 
\end{array}
\right]_A,
\label{eq:transfermat}
\end{eqnarray} 
where $T_{BA}$ is given by 
\begin{eqnarray} 
T_{BA}=
\left[
\begin{array}{cc}
1 & 0 \\
0 & n_A\cos\theta_A/n_B\cos\theta_B
\end{array}
\right].
\label{eq:Tba}
\end{eqnarray} 
Also we define
\begin{eqnarray} 
T_{AB}=T_{BA}^{-1}=
\left[
\begin{array}{cc}
1 & 0 \\
0 & n_B\cos\theta_B/n_A\cos\theta_A
\end{array}
\right].
\label{eq:Tab}
\end{eqnarray} 
The matrices $T_{BA}$ and $T_{AB}$ represent the light propagation
across interfaces $B\leftarrow A$ and $A\leftarrow B$, respectively.
The propagation within one layer is represented by
\begin{eqnarray} 
T_A(d_A)=
\left[
\begin{array}{cc}
\cos\delta_A(d_A) & -\sin\delta_A(d_A) \\
\sin\delta_A(d_A) &  \cos\delta_A(d_A) \\
\end{array}
\right],
\label{eq:Ta}
\end{eqnarray} 
for a layer of type $A$ 
where $d_A$ is the thicknesses of the layer,
and the same expression for $T_B(d_A)$ in which $\delta_A(d_A)$ 
is replaced by $\delta_B(d_B)$.
The phases are given by 
\begin{eqnarray} 
\delta_A(d_A)&=&n_A k d_A/\cos\theta_A,
\nonumber
\end{eqnarray} 
and
\begin{eqnarray} 
\delta_B(d_B)&=&n_B k d_B/\cos\theta_B,
\label{eq:defdelta}
\end{eqnarray} 
where $k$ is the wave number in vacuum.

Now we are ready to consider light propagation
through a Cantor multilayer $C_j$ which is sandwiched
by two media of material of type $A$.
For zero generation layer $A(d_A)$ and 
1st generation layers $A(d_A/3)B(d_A/3)A(d_A/3)$,
the light propagation are respectively given by
\begin{eqnarray} 
M_0&=&T_A(d_A),
\nonumber\\
M_1&=&
T_A\Big(\frac{d_A}{3}\Big)T_{AB}
T_B\Big(\frac{d_A}{3}\Big) T_{BA} T_A\Big(\frac{d_A}{3}\Big).
\label{eq:zerofirst}
\end{eqnarray} 
It can be shown that for $j$-th generation, i.e. $C_j$,
the corresponding matrix $M_j$ is obtained by
recursive replacement of 
\begin{eqnarray} 
T_A\Big(\frac{d_A}{3^{j-1}}\Big),
\label{eq:recursivein}
\end{eqnarray} 
in $M_{j-1}$ by 
\begin{eqnarray} 
T_A\Big(\frac{d_A}{3^j}\Big)T_{AB}
T_B\Big(\frac{d_A}{3^j}\Big) T_{BA} T_A\Big(\frac{d_A}{3^j}\Big).
\label{eq:recursivefin}
\end{eqnarray} 
The transmission coefficient $T$ is given 
in terms of the matrix $M_j$ as
\begin{eqnarray} 
T=\frac{4}{|M_j|^2 +2}
\label{eq:transmission}
\end{eqnarray} 
where $|M_j|^2$ is the sum of the squares 
of the four elements of $M_j$.
This is a quantity measured experimentally 
and has a rich structure with respect to a variation 
of either the wavelength of the light or the number of layers.

Let us consider the simplest experimental setting.
Take the incident light to be normal,
(i.e. $\theta_A=\theta_B=0$)
and also choose the layer $B$ is in vacuum, $n_B=1$.
The results for transmission are shown in Fig.\ref{FIG:transn=2}
for Cantor sequences $C_2$ to $C_6$
where we set $n_A=2$.
The global structure of the transmission coefficient 
apparently show a scaling.
Namely, as the generation of the Cantor sequence is increased,
fine structures appear in addition to the former one. 
We also find a binary structure between complete transmission 
and complete reflection as a function of the wave number
as the generation is increased.
This behavior is distinct from that 
of the Fibonacci case~\cite{REF:KSI}.
The Cantor set under consideration has zero Lebesgue measure 
(in the thermodynamic limit).
This means that, if one points the substrate, 
it is in vacuum, i.e. substrate $B$,
with probability one and the vacuum is dense.
Also there are no isolated points.
The complete reflection is quite striking,
because the above means that if one points the substrate, 
one points the substrate $A$ with probability zero.
Also the complete transmission has to be understood.
So far we do not have any appropriate explanation for
this novel behavior: complete transmission 
and complete reflection.

Acknowledgments: 
This work is supported by the visiting program
of Institute for Solid State Physics, University of Tokyo 
in spring term in 2004.

\end{document}